\documentclass[12pt]{article}
\usepackage[english]{babel}
\usepackage[utf8]{inputenc}
\usepackage{amsmath}
\usepackage{graphicx}
\usepackage{etoolbox}
\usepackage{changepage}
\usepackage{titlesec}
\usepackage[parfill]{parskip}
\usepackage{geometry}
\usepackage{float}
\usepackage[numbers, square, comma, sort&compress]{natbib}
\usepackage[]{notes2bib}

\usepackage{caption}
\usepackage{subcaption}

\usepackage{lipsum} 
\usepackage{soul}
\setstcolor{red}

\usepackage[autostyle]{csquotes} 
\usepackage{graphicx}
\usepackage{graphics}
\usepackage{amsmath}

\usepackage{tikz}
\usepackage{verbatim}
\usetikzlibrary{shapes,shadows,calc}
\usepgflibrary{arrows}

\usepackage{hyperref}

\providecommand{\keyareas}[1]
{
  \textbf{\textit{Conference Key Areas:}} #1
}

\providecommand{\keywords}[1]
{
  \textbf{\textit{Keywords:}} #1
}

\titleformat*{\section}{\normalsize\bfseries} 

\title{\large \textbf{Incorporating microlearning videos, online exercises and assessments into introductory physics}}
\author{\normalsize \textbf{C.\ Netzer}\footnote{Corresponding author, \nolinkurl{c.netzer@tu-berlin.de}}\\
\normalsize Institut f{\"u}r Chemie, Technische Universität Berlin, \\
\normalsize Stra{\ss}e des 17. Juni 115, 10623 Berlin, Germany\\
\vspace{0.5cm}
\normalsize ORCID 0000-0002-4214-9837\\
\normalsize \textbf{A.\ Mittelstädt}\\
\normalsize Institut für Festkörperphysik, Technische Universität Berlin, \\
\normalsize Hardenbergstr. 36, 10623 Berlin, Germany\\
\normalsize ORCID 0000-0002-7587-9385}
\date{} 

\titleformat{\section}
  {\normalsize\normalsize\bfseries}{\thesection}{10pt}{}{}
\titleformat{\subsection}
  {\normalsize\normalsize\bfseries}{\thesubsection}{10pt}{}{}
\titleformat{\subsubsection}
  {\normalsize\normalsize\bfseries}{\thesubsubsection}{10pt}{}{}
\titlespacing{\title}
  {0pt}{0pt}{6pt}
\titlespacing{\section}
  {0pt}{12pt}{6pt}
\titlespacing{\subsection}
  {0pt}{6pt}{6pt}
\titlespacing{\subsubsection}
  {0pt}{6pt}{6pt}

\begin{document}
\maketitle
\thispagestyle{empty}
\pagestyle{empty}


\keyareas{Physics in engineering, Methods, formats and essential elements for online/blended learning}

\keywords{physics, microlearning videos, online exercises, electronic assessments, changes beyond Covid-19}

\section*{ABSTRACT}
We report on the evolution of the largest physics course at the Technical University Berlin during the COVID-19 pandemic, which boosted the development of online learning content, electronic assessments, and, not at least, the reconceptualization of our teaching methods.
By shifting to distance education, we created features that can be easily implemented and combined with any physics-type course at a university level.
Thereby, we incidentally made introductory physics sustainably accessible to students having difficulties visiting the university.
Besides our experiences gained on blended learning methods, we provide a guide to online exercises, microlearning videos, and e-assessments, including exams realized in a virtual setting via the university's Moodle.
Since the course is attended annually by more than 800 students enrolled in at least twelve bachelor degree engineering and all STEM (science, technology, engineering, and mathematics) fields, our reformations greatly impact the student's learning style.
The teaching concepts presented here are supported by results from the Force Concept Inventory, student feedback, and exam results.
The Force Concept Inventory shows that our online teaching did not negatively affect the student's learning process compared with a traditional face-to-face lecture.
Student's feedback shows that the new format and material being available online are received very well.
Finally, the exam results show that virtual exams conducted in a remote setting can be designed to minimize cheating possibilities. In addition, the test was able to yield a similar distribution of points as in the previous traditional ones.

\newpage
\section{INTRODUCTION}
%
Due to the COVID-19 pandemic, from the summer semester 2020 on, we were forced to transform our introductory physics course 
``Physics for Engineers''
into a distance learning format.
We did not have the time to plan and prepare our online teaching methods carefully, but fortunately, we could use the suddenly freed human resources tied up in face-to-face on-site teaching methods.
Therefore, our team was vigorously supported by ten student assistants, many of whom worked overtime to get involved and create online content.
Another advantage was that we started reforms in 2017 and implemented interactive teaching methods such as peer instruction \cite{mazur1997} and new exercises, including concept tests, so we did not start from scratch; cf.\ Ref.\ \cite{mittelstaedt2019}.\\
This paper aims to provide guidance and act as a motivator to implement microlearning videos, online exercises, and virtual exams conducted in a remote setting into any physics-type course; see Sec.\ \ref{sc:methods} for the methods implemented.
Here we share our experiences and suggest instructor-level implications for online teaching methods that can be combined with face-to-face teaching and interactive methods making the teaching approach even more effective.
In particular, we focus on creating material for online learning, laying out a foundation for building interactive formats. 
Also, the effort required to implement the methods and their reception will be discussed.
In Secs.\ \ref{sc:methods2} and \ref{sc:results}, we show our methods for evaluating our new teaching methods and discuss selected results related to the videos and the weekly exercises.
We conducted a Force Concept Inventory (FCI) test \cite{FCIoriginal}, and the exam results show that exams taken by students at home on their own devices can produce comparable results to face-to-face on-site exams.
%
%
%
%
\paragraph{General structure of the course.}
Our physics course is aimed mainly at engineering students in the first two semesters.
For some curricula, this is a compulsory elective for higher semesters also.
Additionally, some students of mathematics, computer sciences, and other fields are attending our module as a part of their extracurricular studies.
The course is entirely organized via the university's Moodle.
It was structured into weekly units consisting of a virtual lecture, non-mandatory online exercises, and additional online material, including videos and lecture notes.\\
To make the course more accessible, we used the lesson activity of Moodle, guiding the students through each week's material.
Here, we organized all the content related to a specific topic in small chunks connected by a golden thread guiding the students.
This became available after each lecture along with the weekly material.
If there was no additional material, participants were first advised to watch the microlearning videos. Then they were asked to read the corresponding chapter in the lecture notes and finally to solve algebraic and conceptual tasks to test their knowledge themselves.
Also, they were reminded each week that they can use our various possibilities to contact us any time.
This golden thread is particularly helpful because the course covers lots of content, beginning with classical mechanics, electromagnetism, and thermodynamics in the first semester and continues with atomic physics, quantum mechanics, nuclear, and solid-state physics during the second.\\
The course is usually passed with a written exam, now changed to a virtual remote setting via Moodle, including automatic grading of most of the questions.
There were no requirements to participate in the exam. However, the students were highly encouraged to use all of the material we offered them, particularly the online exercises.\\
Since the course is mainly attended by first-year students, and because studies suggest that communication among and with students and student's ability to self-organize are essential to the perceived effectiveness of a teaching method \cite{hart2012factors, klein2021studying}, we tried various social media.
Social media are critical to the sustainable success of digital teaching methods, but our experience is discussed elsewhere due to the article's brevity.
After two semesters, the Moodle forum and virtual seminars proved to be successful. Offerings such as chats were hardly used; however, students primarily use messenger channels with peers to communicate with each other.
\section{IMPLEMENTED DISTANCE LEARNING METHODS}\label{sc:methods}
\subsection{Videos}
\begin{figure}[tbh]
    \centering
    \includegraphics[]{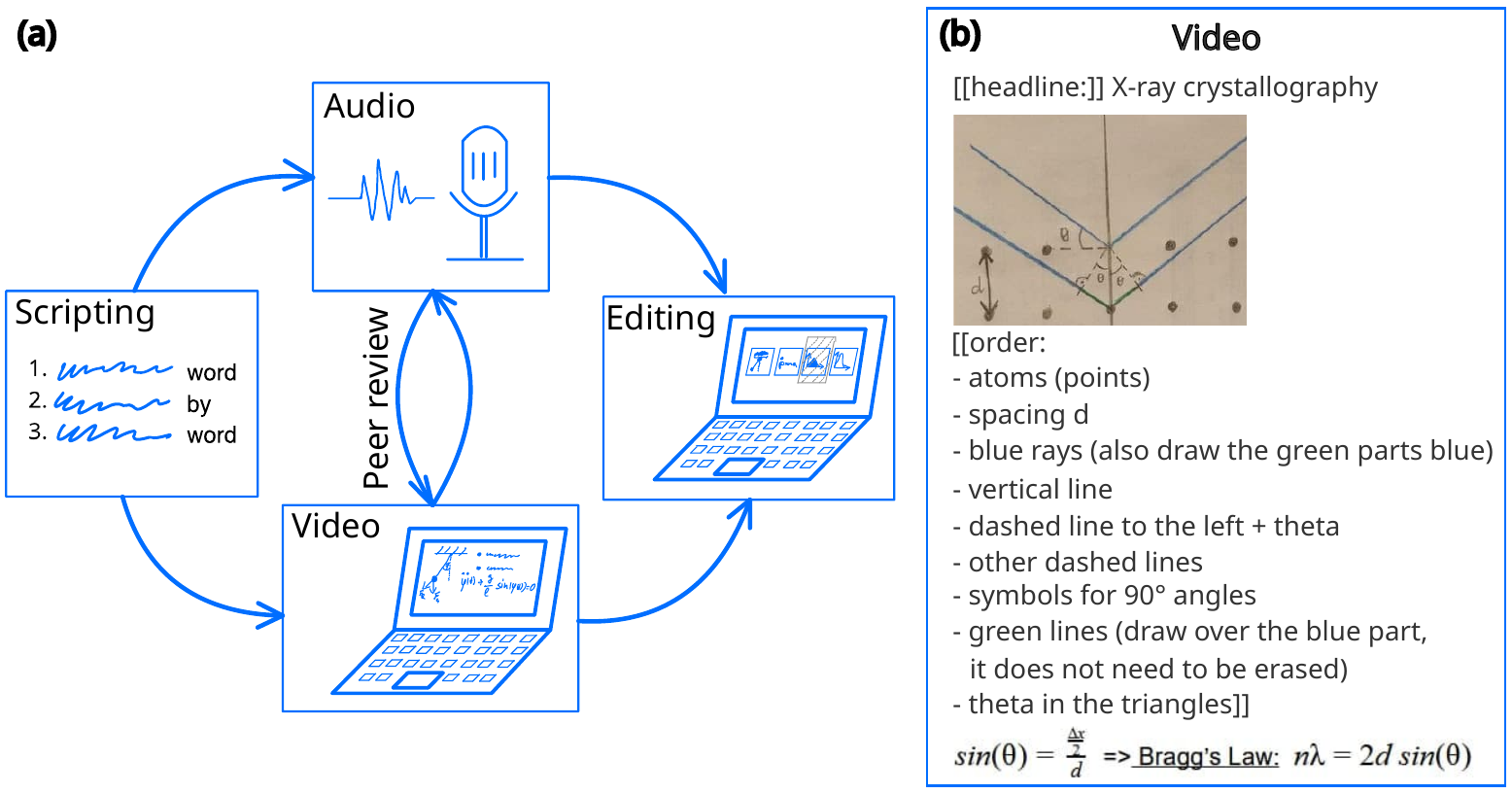} 
    \caption{(a) Flowchart of the process we used to create the videos.
    (b) An example of instructions included in the video recording script.
    }
    \label{fig:videos1}
\end{figure}
While the lecture was not recorded due to the privacy guidelines, we provided short videos covering the essential aspects in just a few minutes.
The creation of the videos was divided into three major steps, visualized in Fig.\ \ref{fig:videos1}(a):
\begin{enumerate}
    \item \textit{Scripting.} Carefully deciding on the text to be spoken and written and the sketches to be drawn was the major accomplishment in improving the videos compared to recordings of a free talk.
    We had the chance not only to avoid filler words and choose efficient and instructive phrases but also to think about possible misunderstandings or inaccuracies that go unnoticed by the speaker when speaking freely, even in a well-planned presentation.\\
    We prepared separate scripts for audio and video, such that one can focus on recording the respective part without being distracted by speaking and writing simultaneously.
    Figure \ref{fig:videos1}(b) shows an example for the video scripts.
    Here a sketch is to be drawn with instructions on the order of its elements, making sure that each new element is added in the identical sequence as in the audio recording.
    %
    There is no need to take the time to work out elaborated sketches or formatting in the scripts as long as the recorder understands how to translate the instructions into the video correctly.
    We also point out the following:
    %
    In the example from Fig.\ \ref{fig:videos1}(b), the script continues with an interlude on wave interference, written on a new sheet.
    In the final edit of the video, this interlude is shown after the sketch is finished.
    Afterward, the video switches back to the screen with the sketch and continues with deriving the formula seen in the example.
    However, this is irrelevant to the process of recording, which in turn shows the flexibility of our video creation process.
    \item \textit{Audio and video recording.} 
    After recording these separately, we double-checked for mistakes and removed the noise from the audio file.
    Another advantage of this way of video production compared with a direct recording of a talk is the possibility to repeat the last sentence if there occurs some mistake in speaking or writing.
    \item \textit{Editing.} 
    Not only are mistakes cut out, but also the pacing can be adjusted when putting audio and video together.
    In particular, the person writing can focus on readable handwriting and carefully draw the sketches without talking while rushing to finish a sketch quickly.
    Of course, the drawing speed of sketches was kept slow enough in cases where a faster pace would be confusing.
\end{enumerate}

We worked in a team of five people by creating the videos, though each could switch with someone for another video. In the following, we outline the workload of creating the videos:\\
The \textit{scripting} is done by (at least) two people.
This is necessary to discuss the wording and the arrangement of sketches, for example.
Thorough preparation for one video of about 15 minutes can take up to one day of scripting, making this the most time-consuming part of the production.\\
We also assigned two people to record \textit{video and audio}, one for each of these tasks. 
Afterward, they both cross-check the recording of the other. 
We found that recording five minutes of final audio, already removing noise and cutting out mistakes, took us about one hour of work.\\
Since no discussion or cross-checking is needed to \textit{edit} the videos, one person is sufficient for this task.
However, we cannot provide a typical amount of time for this, as this highly depends on the contents of the video. For example, some sketches include many changes in color or the use of several tools where a menu bar shows up, which was to be cut out of the video.
See also the note and video at \bibnote[Video]{The interested reader may see the microlearning video at \mbox{(\url{https://youtu.be/UQOHlMiJ6Hs})} or contact the authors to obtain one of these videos as an example.
Note that the videos are in German, but the difference to recorded talks with a slideshow can be seen independently of the language.
}.
\subsection{Exercises}\label{sc:exercises}
In anticipation of the online format of the exam, the exercises were also integrated into Moodle, thus being solved completely online and graded automatically.
Most of our exercises can be sorted into two categories:
\begin{itemize}
    \item \textit{Algebraic exercises with random numbers}.
    Here, the students had to write down a numerical result, including a unit if necessary.
    For example, given the wavelength of a laser pointer and the distance between minima of diffraction caused by a hair in a given distance to the screen, the students had to estimate the hair's breadth, motivating them to try this with their's at home.\\
    While in classical worksheets, points can also be assigned for correct formulas or intermediate steps in miscalculations, Moodle only grades the final result.
    However, this did not turn out to be a problem when the exercise was short enough.
    Also, as the students were allowed and expected to use a cheat sheet and tools such as WolframAlpha \cite{WolframAlpha}, there is no need to grade achievements like ``knowing the correct formula'' or ``computing correctly.''
    Instead, the grading of the final result is considered as grading their derived formula by testing it for the given random values.
    \item
    \textit{Exercises with fixed options}, e.\,g., multiple-choice (MC) questions or drag and drop exercises.
    This type is well suited for conceptual questions and also suitable for algebraic ones with more complicated computations. 
    By giving a few answers to select from, computational errors are readily noticed instead of being graded with zero points.
    If there is an expected typical mistake that should not be given away by this and would not give points in a traditional exercise, it can be included as one of the possible answer options.\\
    We also used this type when the correct answer is somewhat surprising and offered answer options in a more expected range.
    By this, the multiple-choice character does not give away the correct answer, but students are encouraged to select their correct result if found in the options even if they would doubt it at first.
    For example, the time it would take in theory to fall through a hole straight through the earth onto its other side was to be computed as less than an hour while it would take more than half a day to travel around the earth with the speed of sound.\\
    Also, Moodle's ``random short-answer matching'' allows for non-algebraic randomized questions.
    For example, the students had to select the number of translational and rotational degrees of freedom of several gas molecules randomly chosen out of a given list.
\end{itemize}
%
%
%
%
%
%
%
When using algebraic MC questions with random numbers, the following shortcoming is pointed out:
These exercises can be designed in a way that the correct option is either in a fixed position in terms of its numerical value (e.\,g., it is always the third-highest number) or at a random position (e.\,g., using a sine of the random numbers for generating wrong answers).
In the first case, the correct answer is easily communicated when used in an exam, diminishing the effect of using random numbers.
However, in the second case, one of the randomly generated incorrect answers could approximately match the correct one while still being graded as a wrong answer.
To avoid this confusion, one would need to manually check all the rows of numbers generated by Moodle and change them if necessary.\\
%
%
An advantage of the online format is that animated GIFs can be used.
For example, students have to identify different types of damped oscillators from given animations of springs.
Also, animations are helpful to illustrate the scenario in exercises about moving objects.
For example, the rotation of a wire loop relative to a magnetic field is easily represented.\\
The creation of the exercises took about one to two days per week, strongly depending on the exercises.
The most time-consuming factors are the following:
\begin{itemize}
    \item \textit{Developing new questions suitable for the online format}.
    For instance, questions that ask students to explain or draw something or derive a formula need to be adapted accordingly.
    Instead of giving explanations, they had to select the correct reasoning in an MC format, and instead of drawing, they have to select or label the correct diagram. Formulas to be derived were checked by plugging in numbers.
    \item \textit{Creating sketches and animations}. For instance, diagrams are needed when the task was to label them, such as the phase diagram of a Van der Waals gas.
    In other cases, we deemed a sketch of the given scenario helpful rather than explaining geometric details in words.
    %
    %
\end{itemize}
%
%
%
%
%
%
\subsection{Exam}
Since the exam was taken in a virtual setting via Moodle, with each student using their own device at home, two major questions arose:
\begin{itemize}
    \item How to choose appropriate questions for an open-book online exam? 
    \item How to avoid cheating in a remote exam?
\end{itemize}
%
As a general discussion on creating assessment tasks via Moodle is given in Sec.\ \ref{sc:exercises}, we focus on the kind of tasks suitable for a virtual exam.
%
In selecting questions for the exam, we considered the following concept:
\begin{itemize}
    \item 
    \textit{Structure}. 30 to 40\% of the exam questions consist of the weekly exercises, which we announced at the beginning of the semester as motivation to complete the exercises.
    The remaining 60 to 70\% of the exam consists of moderately tricky as well as challenging tasks, allowing for efficient differentiation between grades.
    We weigh the easier of these tasks and the tasks from the weekly exercises with enough points to pass the exam, making it easier to pass the exam by solving the exercises during the semester.
    %
    \item 
    \textit{Concept tasks}. We take care to include only those questions whose answers cannot be easily found on the internet.
    Solutions to some of the conceptual problems are easy to find, but a simple transfer to another situation is often enough to hide the keywords.
    For example, in a paradox of special relativity, a train traveling through a tunnel is replaced by an arrow shot through a pipe.\\
    %
    %
    In the MC-type tasks, we combine possible answers (such as "yes" or "no") with different explanations that include some keywords also used in explanations found on the internet.
    Thus, students who find explanations during the exam must understand them well enough to exclude the wrong answers.
    %
    We also included tasks of the essay question type from Moodle.
    Here we make sure to include at least one easy question where everyone should answer and at least one difficult one where the exact wording should not be the same in any of the answers.
    Copied answers could then be easily found by comparing the student's respective texts.
    \item \textit{Algebraic tasks}. For algebraic tasks where students derive a formula, plug in numbers from the text, and get a numerical result, it is easy to construct examples that avoid finding the answers quickly on the internet.
    To avoid cheating, we use random numbers.
    %
\end{itemize}
Other options avoiding cheating, which we did not use due to time constraints, are: 
Preparing several equivalent questions of a similar type and have one of them randomly drawn.
The order of the questions and the options within the MC questions could also be randomized, making communication between students a bit more complicated, but not very much either.\\
%
%
%
%
Instead, we used a more effective method by including more complex questions than usual.
Thus, we put some pressure on the students to finish the exam in time, not giving them enough time to communicate.
Of course, this strategy alone could also result in a lack of time for solving the question independently.
%
To ensure that the overall difficulty of the exam is not affected, we specified that the best grade is 85\% of the total points instead of 95\%, and that the exam is passed with 40\% instead of 50\%.
This way, students are busy answering all the questions without being very disadvantaged if they did not manage to complete all the questions in time.

\section{EVALUATION METHODS}\label{sc:methods2}
To evaluate the success of our teaching methods and how the students received them, we used the following two methods:\\
First, we evaluated the learning process during the semester.
In addition to the weekly exercises, we included the FCI, which is easily adapted into Moodle's MC-type questions.
The FCI is a test instrument that gives a measure for understanding classical Newtonian mechanics, particularly Newton's axioms and the concept of a force.
The test was offered at the beginning of the semester and after Newtonian mechanics was covered in the lecture.\\
%
%
%
Second, we asked the students for feedback with a mid-term evaluation via Moodle's feedback activity. 
In order to obtain a sufficient amount of answers, we made this activity obligatory for the remaining weekly exercises.
The number of responses we obtained matched the number of participants in the exam quite well, so we consider the results very representative for our audience, despite the student answered anonymously.
%
%
%
%
%
%
%
\section{RESULTS AND DISCUSSION}\label{sc:results}
\begin{figure}[tb]
    \centering
    \includegraphics[]{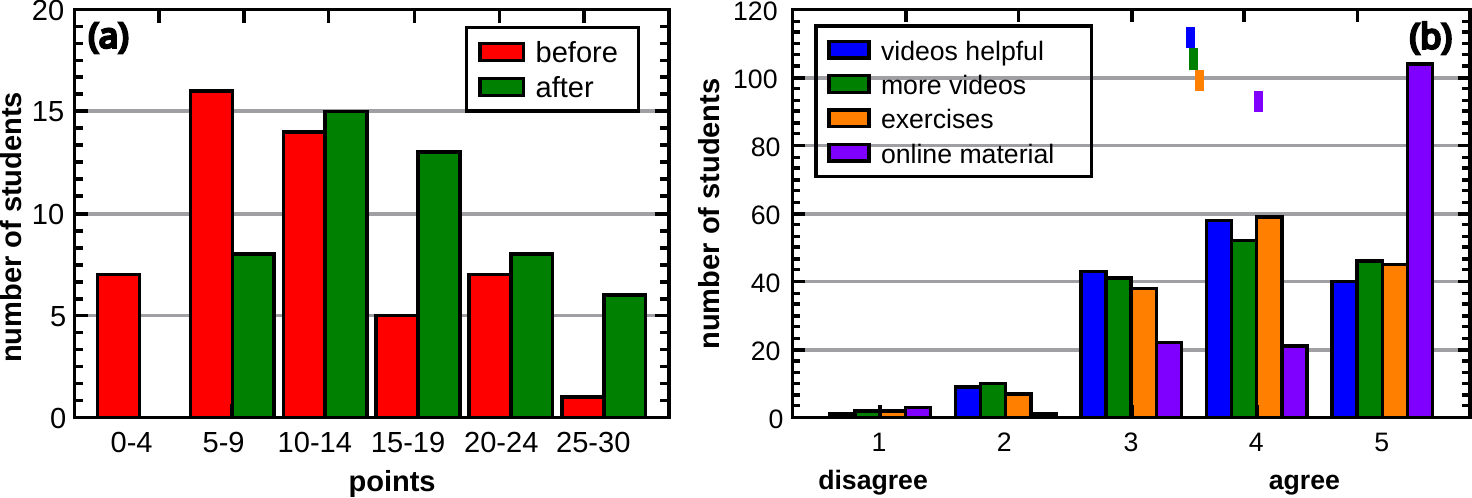}
    \caption{(a) Results of the FCI at the beginning of the semester (``before'', average $11.2$) and after the Newtonian mechanics were discussed in the course (``after'', average $15.7$). A total number of $N=50$ participated in both tests. (b) Answers of students to ``The videos were helpful for my understanding'' (blue), ``There should be more videos in the future'' (green), ``The format of Moodle exercises works well'' (orange), and ``It is helpful that material (exercises, videos, slides, and lecture notes) are available online'' (purple), from 1 for ``absolutely disagree'' to 5 for ``absolutely agree''. Mean values are 3.84 (blue), 3.86 (green), 3.91 (orange), and 4.47 (purple). 
    A total number of $N=151$ students participated in the evaluation.
    }
    \label{fig:eva}
\end{figure}

In this section, we analyze the results obtained from the FCI test, the mid-term evaluation, as well as the exam results and discuss our experiences.\\
%
Figure \ref{fig:eva}(a) shows the results for the students participating in both \textit{FCI} tests.
%
In a meta-analysis by Korff et al.\ \cite{FCImeta} using data from 450 classes, the ``normalized gain'' is used to measure student's improvement.
This is defined as the absolute gain in points divided by the number of wrong answers in the first test.
The meta-analysis found an average normalized gain of $0.22$ for traditional lectures and an average of $0.39$ for courses with interactive engagement.
In our case, the normalized gain was $0.23$.
As seen from our course outline, we mostly offered material for self-learning, making the course more comparable to a traditional lecture.
Thus, we conclude that the online format did not affect the learning progress in any negative way.\\
\begin{figure}[tb]
    \centering
    \includegraphics[width=12cm]{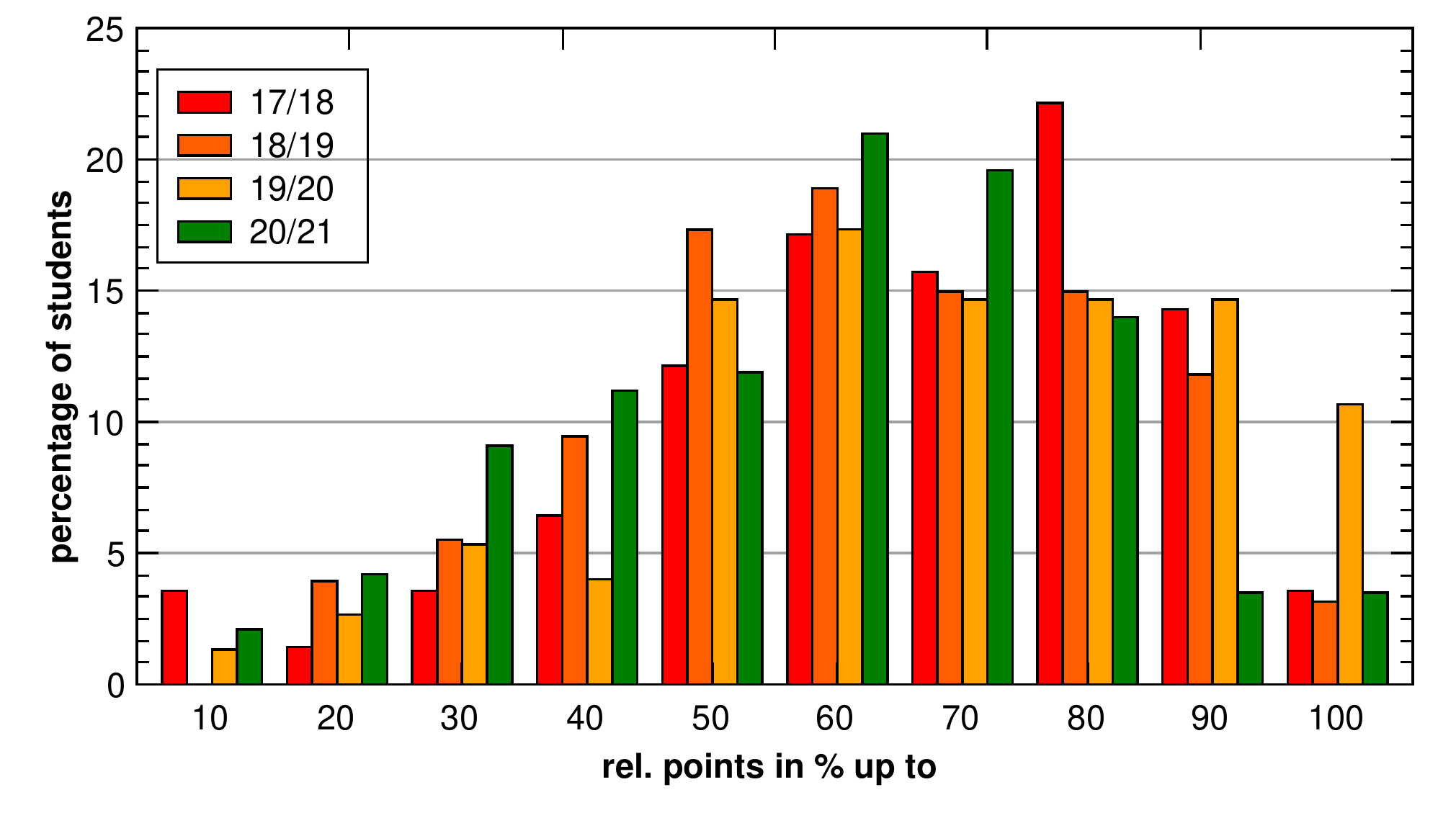}
    \caption{The relative share of students versus relative points obtained in the final online exam at the end of the winter term 2020/21 (green) and the preceding three winter terms (red, orange, yellow). The relative points of the online exam were slightly changed for this graphic to account for the best grade being obtained at a lower number of points.}
    \label{fig:klausuren}
\end{figure}%
%
%
%
%
%
%
%
In the following, we relate the e-learning content discussed above -- videos, exercises, and the exam -- to their respective benefits and evaluate student feedback as much as possible.\\
We start with the \textit{videos}, for which we built a team of five persons as mentioned above, each of them working about one day each week.
Despite this great effort of careful video production, we want to emphasize that the videos will be used in the following years of teaching, which is expected to relieve us from much work in the long term.
This effort was also well received by the students.
In the comments, they complimented on the videos, and in the evaluation, they wished for us to continue producing them; see the green bars in Fig.\ \ref{fig:eva}(b).
Similarly, most of them said the videos were helpful for their learning process; see the blue bars in Fig.\ \ref{fig:eva}(b).\\
%
%
%
As for the videos, the \textit{exercises} will and can be easily reused in upcoming years.
Also, the online exercises are graded automatically, which saved us much time.
We solely needed to hand out a few exercises of the essay question type, which required to be graded manually, as this type was also included in the exam.
Additionally, the automatic grading allowed the students to try the exercises at any time and as many times as they preferred.
We could see that this possibility was frequently used from the timestamps in the attempts.
Also, the students answered in our evaluation that they are happy with the way the exercises are available; see the orange bars in Fig.\ \ref{fig:eva}(b).\\
%
%
%
%
%
%
As the course ended with the \textit{exam}, we did not obtain feedback via an evaluation from the students. Instead, we evaluate the grading.
The grades were distributed in a way we wished for and did not deviate significantly from preceding semesters.
Figure \ref{fig:klausuren} shows the percentage of students obtaining a relative number of points for the online exam and those three years before.
The exam was neither showing a massive amount of best grades, which would be a possible indicator that many students cheated, nor a lot of failed exams, indicating that the online teaching itself failed.
The only notable deviation from previous distributions is the lower percentage of students with 80--90 points, which an unexpectedly tricky question might cause. However, the statistical significance is not clear as there were some outliers on the traditional exams as well, as seen in the 100-point bars.
From these results, we feel well prepared for another online exam.
Although there might have been some students who cheated, it seems that this did not affect the overall grade distribution.
\section{CONCLUSION}
We provide a guide to implementing distance learning formats based on our experiences in the last two semesters. Here we address the production of microlearning videos, online exercises, and the implementation of virtual exam formats.
Our experience shows that online teaching content has been well received by students and evaluated positively. Examinations that are conducted in a virtual and remote setting can be designed in a way that the results are comparable to assessments that are conducted on-site face-to-face.
As shown by the FCI, at least the teaching in Newtonian mechanics did not suffer from the online format than a traditional face-to-face lecture.
The available resources will help and support us with online and face-to-face teaching in the years to come.
Having made some efforts in the first two distance semesters, we have thus laid the foundation for further online teaching while maintaining the quality of our teaching.
As we have focused on creating materials and exercises, we hope to continue to improve our teaching by incorporating them into more interactive formats.
%
%
%
%
%
%
%
%
%
%

\section{ACKNOWLEDGEMENTS}
Special thanks go to the video team with C.\ Fillies, D.\ Schutsch, A.\ Técourt, H.\ Nguyen Van, and J.\ Fialkowski, H.\ Fried, K.\ Hadjikyriakos, F.\ Nippert, J.\ Rausch, M.\ Ries, P.\ Schlaugat, and R.\ Dastgheib Shirazi, without whom this and last year's class of engineering students would have had to live with a significantly reduced teaching program. Furthermore, thanks go to H.\ Grahn, A.\ Schliwa, A.\ Hoffmann, and M.\ Wagner for a new lecturing style, helping creating content and organizing.








\bibliographystyle{unsrtnat} 
\bibliography{bibliography_file}

\begin{thebibliography}{8}
\providecommand{\natexlab}[1]{#1}
\providecommand{\url}[1]{\texttt{#1}}
\expandafter\ifx\csname urlstyle\endcsname\relax
  \providecommand{\doi}[1]{doi: #1}\else
  \providecommand{\doi}{doi: \begingroup \urlstyle{rm}\Url}\fi

\bibitem[Mazur and Hilborn(1997)]{mazur1997}
Eric Mazur and Robert~C. Hilborn.
\newblock Peer instruction: A user's manual.
\newblock \emph{Physics Today}, 50\penalty0 (4):\penalty0 68--69, 1997.
\newblock \doi{10.1063/1.881735}.

\bibitem[Mittelst{\"a}dt and Schliwa(2019)]{mittelstaedt2019}
Alexander Mittelst{\"a}dt and Andrei Schliwa.
\newblock Student activation in curriculum "physics for engineers".
\newblock In \emph{Varietas delectat... Complexity is the new normality:
  Proceedings SEFI 2019, SEFI 47th Annual Conference, Budapest, 16-20
  September, 2019}, pages 787--794, 2019.
\newblock ISBN 978-2-87352-018-2.

\bibitem[Hestenes et~al.(1992)Hestenes, Wells, and Swackhamer]{FCIoriginal}
David Hestenes, Malcolm Wells, and Gregg Swackhamer.
\newblock Force concept inventory.
\newblock \emph{The Physics Teacher}, 30\penalty0 (3):\penalty0 141--158, 1992.
\newblock \doi{10.1119/1.2343497}.

\bibitem[Hart(2012)]{hart2012factors}
Carolyn Hart.
\newblock Factors associated with student persistence in an online program of
  study: A review of the literature.
\newblock \emph{Journal of Interactive Online Learning}, 11\penalty0 (1), 2012.

\bibitem[Klein et~al.(2021)Klein, Ivanjek, Dahlkemper, Jeli{\v{c}}i{\'c},
  Geyer, K{\"u}chemann, and Susac]{klein2021studying}
Pascal Klein, Lana Ivanjek, Merten~Nikolay Dahlkemper, Katarina
  Jeli{\v{c}}i{\'c}, M-A Geyer, Stefan K{\"u}chemann, and Ana Susac.
\newblock Studying physics during the covid-19 pandemic: Student assessments of
  learning achievement, perceived effectiveness of online recitations, and
  online laboratories.
\newblock \emph{Physical Review Physics Education Research}, 17\penalty0
  (1):\penalty0 010117, 2021.
\newblock \doi{10.1103/PhysRevPhysEducRes.17.010117}.

\bibitem[Video()]{Video}
Video.
\newblock The interested reader may see the microlearning video at \mbox {(\url
  {https://youtu.be/UQOHlMiJ6Hs})} or contact the authors to obtain one of
  these videos as an example. Note that the videos are in German, but the
  difference to recorded talks with a slideshow can be seen independently of
  the language.

\bibitem[Inc.()]{WolframAlpha}
Wolfram~Research{,} Inc.
\newblock Mathematica, {V}ersion 12.2.
\newblock URL \url{https://www.wolframalpha.com/}.
\newblock Champaign, IL, 2020.

\bibitem[Von~Korff et~al.(2016)Von~Korff, Archibeque, Gomez, Heckendorf,
  McKagan, Sayre, Schenk, Shepherd, and Sorell]{FCImeta}
Joshua Von~Korff, Benjamin Archibeque, K.~Alison Gomez, Tyrel Heckendorf,
  Sarah~B. McKagan, Eleanor~C. Sayre, Edward~W. Schenk, Chase Shepherd, and
  Lane Sorell.
\newblock Secondary analysis of teaching methods in introductory physics: A
  50k-student study.
\newblock \emph{American Journal of Physics}, 84\penalty0 (12):\penalty0
  969--974, 2016.
\newblock \doi{10.1119/1.4964354}.

\end{thebibliography}


\end{document}